\documentclass[prd,onecolumn,notitlepage,amsmath,nofootinbib,superscriptaddress,showkeys]{revtex4-1}

\usepackage[T1]{fontenc}
\usepackage[latin9]{inputenc}
\usepackage[english]{babel}

\usepackage{amsmath}
\usepackage{graphicx}
\usepackage{amssymb}
\usepackage{esint}
\usepackage[usenames,dvipsnames]{color}

\usepackage{longtable}
\usepackage{dcolumn}
\usepackage{ulem}
\usepackage{nicefrac}

\makeatother

\usepackage{babel}

\makeatletter

\begin{document}

\title{A Second Quantized Approach to the Rabi Problem}

\author{M. C. Baldiotti}
\email{baldiotti@uel.br}
\affiliation{
Departamento de F\'{\i}sica, Universidade Estadual de Londrina, 86051-990,
Londrina, Paran\'{a}, Brazil
}

\author{C. Molina}
\email{cmolina@usp.br}
\affiliation{Escola de Artes, Ci\^{e}ncias e Humanidades, Universidade de S\~{a}o Paulo \\
Av. Arlindo Bettio 1000, CEP 03828-000, S\~{a}o Paulo-SP, Brazil}

\begin{abstract}
In the present work the Rabi Problem, involving the response of a spin~$\nicefrac{1}{2}$ particle subjected to a magnetic field, is considered in a second quantized approach. In this concrete physical scenario, we show that the second quantization procedure can be applied directly in a non-covariant theory. The proposed development explicits not only the relation between the full quantum treatment of the problem and the semiclassical Rabi model, but also the connection of these approaches with the Jaynes-Cummings model. The consistency of the method is checked in the semiclassical limit. The treatment is then extended to the matter component of the Rabi problem so that the Schr\"{o}dinger equation is directly quantized. Considering the spinorial field, the appearance of a negative energy sector implies a specific identification between Schr\"{o}dinger's and Maxwell's theories. The generalized theory is consistent, strictly quantum and non-relativistic.
\end{abstract}


\maketitle

\section{\label{Introduction}Introduction}

Two level systems are paramount in Quantum Mechanics. From the purely
theoretical point of view, they are the simplest and best understood quantum
system. From the phenomenological side, they adequately describe important
physical scenarios, that include lasers \cite{Nus73}, optical resonance
\cite{AllEb75}, resonance absorption \cite{DreKe61}, nuclear induction
experiments \cite{RabRaS54}. Recently two state quantum systems modeling atoms
and molecules in cavities subjected to electromagnetic fields \cite{FeyVe57}
have been receiving attention due to their relation with possible
implementations of quantum computers (see for example \cite{LosDi98,bal2}).

In the semiclassical approach, that is when external fields are considered
classically, to treat a two level problem is to solve a spin equation with the
form \cite{BagGiBL05}
\begin{equation}
i\frac{d\psi}{dt}=\hat{H}\psi\,,\;\hat{H}=\left(  \mathbf{F}\boldsymbol{\sigma
}\right)  ~, \label{a.3}%
\end{equation}
where $\mathbf{F}=\left(  F_{1}\left(  t\right)  ,F_{2}\left(  t\right)
,F_{3}\left(  t\right)  \right)  $ is an arbitrary vector field which depends
on time, and $\boldsymbol{\sigma}=\left(  \sigma_{1},\sigma_{2},\sigma
_{3}\right)  $ denotes the Pauli matrices.

An important particular case of a two state quantum system is the Rabi
Problem, a spin~$\nicefrac{1}{2}$ particle subjected to a constant magnetic
field orthogonal to a second rotating field \cite{RabRaS54}. The Rabi problem
can be dealt with Eq.~(\ref{a.3}) when the external field has the form
$\mathbf{F}=\left(  B\cos\omega t,B\sin\omega t,B_{3}\right)  $, where $B$,
$\omega$ and $B_{3}$ are real constants. Another important problem, related to
the Rabi case, is obtained with a linearly polarized field, $\mathbf{F}%
=\left(  2B\cos\omega t,0,B_{3}\right)  $. This latter system does not have
exact solution, but can be adequately described by a decomposition in the
associated Hamiltonian in the form
\begin{equation}
\hat{H}=\hat{H}_{+}+\hat{H}_{-}~,\ \hat{H}_{\pm}=\left(  \mathbf{B}_{\pm
}\boldsymbol{\sigma}\right)  ~,\ \mathbf{B}_{\pm}=\left(  B\cos\omega t,\pm
B\sin\omega t,B_{3}\right)  ~.\label{hl0}%
\end{equation}
The main point in this decomposition is that, near the resonance, for $\omega
B_{3}>0$, the term connected to the component $\mathbf{B}_{-}$ does not
contribute to the transitions. In this way the linearly polarized field
problem is reduced to the Rabi problem. The quantity $\mathbf{B}_{-}$ is the
so-called counter-rotating term, and the discarding of this term is usually
referred as the rotating wave approximation \cite{Sten73}.

The interaction of a two level system with an quantized electromagnetic field
was presented in the work of Jaynes-Cummings \cite{jc}. For the case of a
lossless cavity with a single mode and a uniform field, the interaction
Hamiltonian of the Jaynes-Cummings model is given by
\begin{equation}
\hat{H}_{\mathrm{int}}=\Omega\hat{B}\hat{S}~,\ \hat{B}=\hat{b}+\hat{b}%
^{+}~,\ \hat{S}=\sigma_{1}~.\label{jc}%
\end{equation}
In this expression $\hat{B}$ is the field operator, written in terms of the
creation and annihilation operators ($\hat{b}^{+}$ and $\hat{b}$) of the
electromagnetic field. The operator $\hat{S}$ is the spin operator, that is,
the polarization operator of the two level system.

The original Rabi Hamiltonian \cite{RabRaS54} presents a semiclassical
approach involving a circularly polarized field. The Jaynes-Cumming model
involves a second quantized treatment of a linearly polarized field 
and cannot be obtained
from the Rabi setup without approximations \cite{Yimim}. Although the
connection between Rabi and Jaynes-Cummings models were already investigated
in \cite{jc}, the theme is far from exhausted. For instance, in a recent
development \cite{Yimim}, a new approximation (the so-called \textquotedblleft
intermediate rotating wave approximation\textquotedblright) is introduced.
Based on this approach, the authors relate the Jaynes-Cummings and Rabi
models, even for large values of detuning and coupling. However, the quantum
``Rabi's Hamiltonian'' treated in
\cite{Yimim} is not the Hamiltonian that is obtained\footnote{In many
references, the \textquotedblleft Rabi model\textquotedblright\ is understood
as a two level system coupled with a quantum harmonic oscillator. That
approach does not constitute a second quantization of the original Rabi's
setup \cite{RabRaS54,Rab37} that describes the circularly polarized field. For
example, the Hamiltonian presented in Eq.~(1) of \cite{Yimim} actually
represents the interaction of a spin~$\nicefrac{1}{2}$ particle with a
quantized harmonic oscillator.} from the (second) quantized version of the
problem introduced in Rabi's original developments \cite{RabRaS54,Rab37}. As
one of the contributions of present work, we will give a new view of this relation.

The usual methods for second quantization \cite{berezin} can only be directly
applicable to relativistic and covariant (or non-interacting) theories
\cite{wald,CouWr2016,Bar54,Araki}. However, as it will be demonstrated here,
in some cases this restriction can be relaxed if proper caution is exercised.
Besides, non-equivalence between representations turns the definition of a
semi-classical limit for a second-quantized theory into a highly non-trivial
problem \cite{Ume82}. Therefore, the consistency of the semi-classical limit
must be tested case-by-case. We will see in the specific Rabi's scenario that
the quantization with Schr\"{o}dinger and Heisenberg picture generate distinct
and non-equivalent theories.

Considering the issues discussed, we will present a consistent method for the
second quantization of the (non-covariant and interacting) Rabi problem. This
quantization is made without the rotating wave approximation, and therefore
our development is not equivalent to the Jaynes-Cumming approach. This will be
the main contribution of the present work. In addition, we will show how the
method can be extended to the treatment of a linear field, analyzing the
connection between the Rabi model and the semiclassical limit of the
Jaynes-Cummings approach.

The structure of this work is commented in the following. In
section~\ref{Rabi-problem}, the semiclassical Rabi problem is reviewed. In
section~\ref{quantization} we present our main results, introducing a second
quantized approach for the Rabi setup. Also in this section the semiclassical
limit of the model is discussed, confirming the consistency of the
quantization method. The connection between our development and the original
Jaynes-Cummings is commented. In section~\ref{spinorial-field}, we extend our
treatment to include the spinorial field that describes the
spin~$\nicefrac{1}{2}$ particle of the Rabi setup. Final considerations are
presented in section~\ref{final-comments}.

\section{\label{Rabi-problem}The semiclassical Rabi problem}

For a spin~$\nicefrac{1}{2}$ particle with mass $m$, $g$-factor, electric
charge $q$ and magnetic moment $\mu$, confined in a region with a magnetic field
$\mathbf{B}$, the semiclassical description is given by a Hamiltonian with the
Stern-Gerlach term,
\begin{equation}
\hat{H}=-\mu\left(  \mathbf{B}\boldsymbol{\sigma}\right)  ~,\ \mu=g\frac
{q}{2m}~. \label{Stern-Gerlach}%
\end{equation}
In Eq.~(\ref{Stern-Gerlach}) the field $\mathbf{B}$ is determined by functions
that represent the classical field being considered. We will set $\mu=-1$ to
simplify notation. We are interested in Pauli's theory for a circular field,
i.e., a field in the form
\begin{equation}
\mathbf{B}=\left(  B\cos\omega t,B\sin\omega t,B_{3}\right)  ~. \label{circ1}%
\end{equation}
For this configuration, we have the Rabi's Hamiltonian \cite{Rab37}
\begin{equation}
\hat{H}_{r}=\sigma_{3}B_{3}+B\left(  \sigma_{1}\cos\omega t+\sigma_{2}%
\sin\omega t\right)  ~. \label{rabi}%
\end{equation}

The problem can be elegantly dealt with the use of a rotating frame
\cite{RabRaS54,FeyVe57}. This non-inertial reference system rotates with the
field and hence the field is constant according to rotating observers. The
passage to the rotating frame is done with the unitary transformation
\begin{equation}
\hat{R}_{z}\left(  \omega t\right)  =\exp\left(  -i\frac{\omega}{2}\sigma
_{3}t\right)  ~. \label{rot}%
\end{equation}
The Hamiltonian $\hat{H}^{\prime}$, which describes the problem in the
rotating coordinate system, has the form
\begin{equation}
\hat{H}^{\prime}=\hat{R}_{z}^{+}\hat{H}_{r}\hat{R}_{z}-i\hat{R}_{z}^{+}%
\frac{\partial\hat{R}_{z}}{\partial t}=\frac{\delta}{2}\sigma_{3}+\sigma
_{1}B~, \label{hlinha}%
\end{equation}
where $\delta=2B_{3}-\omega$ is the so-called detuning parameter. The
semiclassical problem has exact solution, and the spin transition is given by
\cite{bal}
\begin{equation}
\left\vert \left\langle +\right\vert U^{\prime}\left(  t\right)  \left\vert
-\right\rangle \right\vert ^{2}=\frac{B^{2}}{\tilde{\Omega}^{2}}\sin
^{2}\left(  \tilde{\Omega}t\right)  ~,\ \sigma_{3}\left\vert \pm\right\rangle
=\pm\left\vert \pm\right\rangle ~, \label{trans}%
\end{equation}
with $\tilde{\Omega}^{2}=\left(  \delta/2\right)  ^{2}+B^{2}$. The quantity
$\tilde{\Omega}$ is denoted as Rabi's frequency. The transition probability in
Eq.~(\ref{trans}) has a maximum for $B_{3}=\omega/2$, and the resonant
frequency of the system is $\omega_{R}=2B_{3}$.

\section{\label{quantization}Quantization of the Rabi problem}

\subsection{A second quantized approach}

Since we are interested in the problem of a particle with spin, fixed in space
and interacting with a magnetic field, we can ignore the electric field and
the spatial dependence of the field, and hence its associated Hamiltonian
density is given by
\begin{equation}
H_{f}=\frac{1}{2}\left\vert \mathbf{B}\right\vert ^{2}=\frac{1}{2}\left(
\left\vert B_{1}\right\vert ^{2}+\left\vert B_{2}\right\vert ^{2}+\left\vert
B_{3}\right\vert ^{2}\right)  ~.
\end{equation}
We will consider $B_{3}=0$. Introducing the quantity
\begin{equation}
b=\sqrt{\frac{1}{2\alpha}}\left(  B_{1}-iB_{2}\right)  ~, \label{b}%
\end{equation}
where $\alpha$ is a positive real constant, the Hamiltonian assumes the form
\begin{equation}
H_{f}=\alpha b^{\ast}b~. \label{hc}%
\end{equation}

The quantization is implemented promoting $b$ to an operator $\hat{b}$, and
imposing the commutation rule
\begin{equation}
\left[  \hat{b},\hat{b}^{+}\right]  =1\Rightarrow\left[  \hat{H}_{f},\hat
{b}\right]  =-\alpha\hat{b}~. \label{rc}%
\end{equation}
The constant $\alpha$ is fixed if we impose that the dynamics is determined by
the Heisenberg equation
\begin{equation}
\frac{d\hat{b}}{dt}=i\left[  \hat{H}_{f},\hat{b}\right]  +U^{+}\frac
{\partial\hat{b}_{S}}{\partial t}U=\left.  \left(  \frac{db}{dt}%
+\frac{\partial b}{\partial t}\right)  \right\vert _{b\rightarrow\hat{b}}~,
\label{emh}%
\end{equation}
where $\hat{b}_{S}$ is the $\hat{b}$ operator in the Schr\"{o}dinger
representation. In the last equality in Eq.~(\ref{emh}), the notation used
means that we first calculate the classical quantities into the parenthesis,
afterwards promoting them to operators. The previous requirement, summarized
in Eq.~(\ref{emh}), can be interpreted as a correspondence principle for the
second quantization approach, ensuring that the quantization has the correct
semiclassical limit.

We are interested in the circular magnetic field presented in Eq.~(\ref{circ1}),
whose intensity is conserved. For this field we have
\begin{equation}
b\left(  t\right)  =\frac{B}{\sqrt{2\alpha}}\exp\left(  -i\omega t\right)  ~,
\label{1}%
\end{equation}
implying that
\begin{equation}
\dot{b}=-i\omega b\,.
\end{equation}
In Eq.~(\ref{1}) it was explicitly considered that $\omega>0$. For $\omega<0$
we should redefine $b$ with the transformation $B_{2}\rightarrow-B_{2}$. As we
will see, this latter prescription is necessary in order that $\alpha>0$. In
the second quantization scheme, the field $\mathbf{B}$ is decomposed in a
basis of orthogonal functions $\phi_{i}$,
\begin{equation}
\mathbf{B}=\sum_{i}\mathbf{c}_{i}\left(  t\right)  \phi_{i}\left(
\mathbf{x}\right)  ~,\ \mathbf{c}_{i}=c_{i}\mathbf{\hat{e}}_{i}%
~,\ \left\langle \phi_{i}\right\vert \left.  \phi_{j}\right\rangle
=\delta_{ij}~. \label{decomp}%
\end{equation}
The particular basis of functions to be chosen is determined by boundary
conditions of the classical problem and symmetries of the system. Besides,
since $\mathbf{B}$ obeys the wave equation (vacuum Maxwell's equations), the
expansion coefficients $\{\mathbf{c}_{i}\}$ obey the classical relation
\cite{sakua}
\begin{equation}
\ddot{c}_{i}=-\omega^{2}c_{i}\Rightarrow\dot{c}_{i}=\pm i\omega c_{i}%
\ ,\ \omega\in\mathbb{R}~. \label{c}%
\end{equation}
In Eq.~(\ref{c}) we used the positivity of the Hermitian operator $\left(
i\boldsymbol{\nabla}\right)  ^{2}$, with any appropriate boundary condition
that makes it a self-adjoint operator.

In the general case, the coefficients $c_{i}$ that appear in Eq.~(\ref{c}) are
linear combinations of the coefficients $b_{i}$ present in the
Hamiltonian~(\ref{hc}). But in the special case of a circular fields,
expressions~(\ref{c}) and (\ref{1}) show that we can identify $c=b$. We stress
that this result is a consequence of the fact that, for the circular field,
the terms $b^{\ast}$ and $b$ that appear in the Hamiltonian in Eq.~(\ref{hc})
are automatically the field expansion coefficients. That implies the choice of
periodic boundary conditions, and also establishes the physical interpretation
of the creation and annihilation operators $\hat{b}^{+}$ and $\hat{b}$. As we
will see, $c=b$ sets that $\hat{b}^{+}$ creates ($\hat{b}$ annihilates)
photons with circular polarization.

We now need to determine the temporal dependence of $\hat{b}_{S}$. In analogy
to the semiclassical treatment, we have the option of considering the field
temporal dependence both the Heisenberg and Schr\"{o}dinger picture. However,
the physical meaning of the field is different when comparing with the simpler
approach in section \ref{Rabi-problem}. This happens because the classical
fields to be quantized must be described in an inertial frame. In addition,
once the dynamic is given by Eq.~(\ref{emh}), the Heisenberg representation
must be associated with this frame. It turns out that we have no freedom in
the association of the representation and the frame that describe the
semi-classical theory. In fact, we could try to implement the temporal
dependence of the circular field in the Schr\"{o}dinger picture, i.e., make
$\partial\hat{b}_{S}/\partial t=-i\omega\hat{b}_{S}$. But using Eqs.~(\ref{1})
and (\ref{rc}), we have from Eq.~(\ref{emh}) that
\begin{equation}
\frac{d\hat{b}}{dt}=-i\alpha\hat{b}+U^{+}\frac{\partial\hat{b}_{S}}{\partial
t}U=-i\omega\hat{b}~. \label{h2}%
\end{equation}
From Eq.~(\ref{h2}), insisting with the Schr\"{o}dinger picture, we would
obtain
\begin{equation}
-i\alpha\hat{b}=-i\omega\left(  \hat{b}-\hat{b}\right)  =0\Rightarrow
\alpha=0~,
\end{equation}
and we could not proceed with the quantization process.

Therefore, the choice of the original frame (classically an inertial frame) is
crucial to the second quantization of a non-covariant theory. Physically this
is reminiscent of the fact that a static field, i.e., a field which does not
depend on time in an inertial frame, does not produce radiation. That is why
we do not try to include a non-null $B_{3}$ in the above procedure.

As a consequence, the temporal dependence must be implemented with the
Heisenberg representation and the field operators do not depend on time in the
Schr\"{o}dinger representation, that is $\partial\hat{b}_{S}/\partial t=0$.
Therefore Eq.~(\ref{h2}) implies
\begin{equation}
i\alpha\hat{b}=i\omega\hat{b}\Rightarrow\alpha=\omega\Rightarrow\hat{H}%
_{f}=\omega\hat{N}~,\ \hat{N}=\hat{b}^{+}\hat{b}~. \label{hamil}%
\end{equation}
Besides, since in all quantization schemes (in flat spacetime) the time enters
as a parameter \cite{Gal02}, we rewrite Eq.~(\ref{1}) as
\begin{equation}
\hat{b}\left(  t\right)  =\hat{b}\,\exp\left(  -i\omega t\right)  ~,\ \hat
{b}\equiv\hat{b}\left(  t=0\right)  ~. \label{2}%
\end{equation}
Since in $t=0$ we have a magnetic field pointing in the $\hat{x}$ direction,
the operator $\hat{b}$ (which do not depend on time) also describes our system
in a reference frame that rotates with the field.

It should be observed that $\hat{N}$ does not depend on time in both
representations. The fact that it is possible to define a single number
operator which is time independent is a consequence of the fact that the
intensity of the classical field $\mathbf{B}$ is constant. Moreover, both
$\hat{b}\left(  t\right)  $ and $\hat{b}$ respect the same algebra~(\ref{rc}).
This implies that the quantum theories constructed with these operators,
although different, are unitarily equivalent.

To complete the description of the physical system discussed here, we
introduce the circular polarization unit vector $\boldsymbol{\epsilon}$ and,
inverting (\ref{b}), we write
\begin{equation}
\mathbf{B}=B_{1}\mathbf{\hat{e}}_{1}+B_{2}\mathbf{\hat{e}}_{2}=\sqrt{\omega
}\left[  b\boldsymbol{\epsilon}+\left(  b\boldsymbol{\epsilon}\right)  ^{\ast
}\right]  ~,\ \boldsymbol{\epsilon}=\frac{1}{\sqrt{2}}\left(  \mathbf{\hat{e}%
}_{1}-i~\mathbf{\hat{e}}_{2}\right)  ~. \label{pol}%
\end{equation}
That is, $b$ and $b^{\ast}$ are the coefficients of the field $\mathbf{B}$ in
the circular polarization basis. It follows, from Eqs.~(\ref{1}) and
(\ref{pol}), that $\hat{b}^{+}$ and $\hat{b}$ create and annihilate photons
with polarization $\boldsymbol{\epsilon}^{\ast}$ and energy $\omega$. For
$\omega<0$ the factors in Eq.~(\ref{pol}) interchange.

With a full description for the free field sector of the model, we can
reintroduce the spin and return to Rabi's Hamiltonian. In Dirac notation,
using~(\ref{2}), Eq.~(\ref{rabi}) can be written as
\begin{equation}
\hat{H}_{r}=\Omega\left(  e^{-i\omega t}\hat{b}\left\vert +\right\rangle
\left\langle -\right\vert +e^{i\omega t}\hat{b}^{+}\left\vert -\right\rangle
\left\langle +\right\vert \right)  +B_{3}\sigma_{3}~,
\end{equation}
where $\Omega\left(  m,q,\omega\right)  \equiv-\mu\sqrt{2\omega}$. The total
Hamiltonian $\hat{H}_{f}+\hat{H}_{r}$, obtained after the second quantization
process, is then
\begin{gather}
\hat{H}=\hat{H}_{0}+\hat{H}_{I}~,\ \hat{H}_{0}=\omega\hat{N}+B_{3}\sigma
_{3}~,\nonumber\\
\hat{H}_{I}=\Omega e^{-i\omega t}\hat{b}\left\vert +\right\rangle \left\langle
-\right\vert +\Omega e^{i\omega t}\hat{b}^{+}\left\vert -\right\rangle
\left\langle +\right\vert ~. \label{h}%
\end{gather}
Notice that Eq.~(\ref{h}) is a result of the formalism, obtained from the
quantization of the Rabi's Hamiltonian in Eq.~(\ref{rabi}). No \textit{ad hoc}
terms were added to Eq.~(\ref{rabi}) for the derivation of Eq.~(\ref{h}).

It should be observed that for any self-state $\left\vert n\right\rangle $ of
the operator $\hat{N}$ we have
\begin{equation}
|\Omega\hat{b}|^{2}\left\vert n\right\rangle =|\Omega|^{2}\hat{b}^{+}\hat
{b}\left\vert n\right\rangle =\Omega^{2}n\left\vert n\right\rangle
=|\mathbf{\hat{B}}|^{2}\left\vert n\right\rangle ~,
\end{equation}
where the normal ordering is chosen for $|\hat{b}|^{2}$. Hence, the classical
field intensity is proportional to the number of photons and their coupling to
the particle,
\begin{equation}
n\Omega^{2}=\left\vert \mathbf{B}\right\vert ^{2}~. \label{omega}%
\end{equation}

Let us now consider the Schr\"{o}dinger representation, where
\begin{gather}
\hat{H}_{S}=\hat{H}\left(  t=0\right)  =\hat{H}_{0}+\hat{H}_{SI}~,\nonumber\\
\hat{H}_{SI}=\Omega\left(  \hat{b}\left\vert +\right\rangle \left\langle
-\right\vert +\hat{b}^{+}\left\vert -\right\rangle \left\langle +\right\vert
\right)  ~.\label{hs}%
\end{gather}
We observe that (\ref{hs}) can also be obtained if we apply a
rotation~(\ref{rot}) in relation~(\ref{h}). This shows that both
Schr\"{o}dinger and Heisenberg representations are connected by a rotation.
But in the second quantization approach used in this work, the rotation
operation no longer has its original meaning. Although the observer
associated with the rotating frame detects a constant field, this same
observer also detects a non-zero number of photons. This remark would not be
true if this observer tries to quantize this constant field. In conclusion,
both frames are not physically equivalent with the second quantization
process.

\subsection{\label{sce}Semiclassical limit of the electromagnetic field}

A well-defined semiclassical limit is important to guarantee the consistency
of the theory. We will investigate this limit presently. In special, we will
see that in our development the semiclassical limit is not associated neither
to the Schr\"{o}dinger nor to the Heisenberg representation. The interaction
representation is the relevant one.

Considering the Hamiltonian in Eq.~(\ref{hs}), we can write \cite{Sten73}
\begin{equation}
\hat{H}_{S}=\hat{H}_{0}^{\prime}+\hat{H}_{1}~, \label{h3}%
\end{equation}
with
\begin{equation}
\hat{H}_{0}^{\prime}=\omega\left(  \hat{N}+\frac{\sigma_{3}}{2}\right)
~,\ \hat{H}_{1}=\frac{\delta}{2}\sigma_{3}+\hat{H}_{SI}~.
\end{equation}
The important point concerning Eq.~(\ref{h3}) is that
\begin{equation}
\left[  \hat{H}_{0}^{\prime},\hat{H}_{1}\right]  =0~, \label{comut}%
\end{equation}
and therefore we can use an interaction picture where the Hamiltonian $\hat
{H}_{0}^{\prime}$ can be eliminated and the dynamics is given by the evolution
operator $U_{1}$, with
\begin{equation}
U_{1}=e^{i\hat{H}_{0}^{\prime}t}e^{-i\hat{H}_{S}t}=\exp\left(  -i\hat{H}%
_{1}t\right)  ~. \label{u1}%
\end{equation}

From relation~(\ref{comut}) we have
\begin{equation}
\exp\left(  i\hat{H}_{0}^{\prime}t\right)  \hat{H}_{1}\exp\left(  -i\hat
{H}_{0}^{\prime}t\right)  =\hat{H}_{1}~,
\end{equation}
implying that
\begin{equation}
e^{i\omega\hat{N}t}\hat{b}e^{-i\omega\hat{N}t}=e^{-i\omega t}\hat{b}~.
\end{equation}
That is, $\exp(i\omega\hat{N}t)$ is equivalent to a rotation $\hat{R}%
_{z}\left(  \omega t\right)  $. Consequently, the transformation that links
the interaction and Schr\"{o}dinger representations,
\begin{equation}
\exp\left(  i\hat{H}_{0}^{\prime}t\right)  =e^{i\omega\hat{N}t}\hat{R}_{z}%
^{+}\left(  \omega t\right)  ~, \label{trans_link}%
\end{equation}
do not alter the frame (the rotating frame). This happens because
$e^{i\omega\hat{N}t}$ and $\hat{R}_{z}^{+}\left(  \omega t\right)  $ represent
rotations in opposite directions. But the transformation~(\ref{trans_link})
discounts the energy associated to the Hamiltonian $\hat{H}_{0}^{\prime}$. In
this way, the vacuum energy in the interaction picture is redefined so that
the energy of the field is not considered.

As will be seen in the following, the interaction frame gives the
semiclassical description. First, by noting that $\left\vert n,-\right\rangle
$ is a eigenstate of $\hat{H}_{1}^{2}$, with eigenvalues $\tilde{\omega}%
^{2}=\left(  \delta/2\right)  ^{2}+n\Omega^{2}$, one can determine the
transition probability
\begin{equation}
\left\vert \left\langle +,n-1\right\vert e^{-i\hat{H}_{1}t}\left\vert
n,-\right\rangle \right\vert ^{2}=\frac{n\Omega^{2}}{\tilde{\omega}^{2}}%
\sin^{2}\left(  \tilde{\omega}t\right)  ~,
\end{equation}
which, from Eq.~(\ref{omega}), we can identify with (\ref{trans}). Another
consequence of decomposition in Eq.~(\ref{h3}) is that, for any value of $n$,
the Hamiltonian $\hat{H}_{0}^{\prime}$ is degenerate for the states $\left(
\left\vert n,\pm\right\rangle ,\left\vert n\pm1,\mp\right\rangle \right)  $.
Moreover, the Hamiltonian $\hat{H}_{1}$ has non-null matrix elements only for
transitions between the same states, or for transitions in the form
$\left\vert n,\pm\right\rangle \rightarrow\left\vert n\pm1,\mp\right\rangle $.
In this way, if the base elements $\left\{  \left\vert n,\pm\right\rangle
\right\}  $ can be reorganized as $\left\{  \left\vert n,+\right\rangle
,\left\vert n+1,-\right\rangle \right\}  $, the Hamiltonian~(\ref{h3}) assumes
a block diagonal form, where each block in position $n$ is a matrix $2\times
2$, which we will denote as $\hat{H}^{\left(  n\right)  }$
\begin{equation}
\hat{H}^{\left(  n\right)  }=\hat{H}_{0}^{\prime\left(  n\right)  }+\hat
{H}_{1}^{\left(  n\right)  }~,
\end{equation}
with
\begin{equation}
\hat{H}_{0}^{\prime\left(  n\right)  }=\omega\left(  n+\frac{1}{2}\right)
\mathbb{I}~,\ \hat{H}_{1}^{\left(  n\right)  }=\frac{\delta}{2}\sigma
_{3}+\Omega\sqrt{n+1}\sigma_{1}~. \label{H0linha_n}%
\end{equation}

The energy associated to the Hamiltonian $\hat{H}_{0}^{\prime\left(  n\right)
}$, which is discounted in the interaction representation, do not correspond
to the magnetic field energy only. The term $\omega/2$ of the constant
multiplying the unit matrix $\mathbb{I}$ in $\hat{H}_{0}^{\prime\left(
n\right)  }$ is originated from the operator $\sigma_{3}/2$, and it is
associated to a photon that the particle can emit\ to the cavity. This energy
will also be discounted in the semiclassical Hamiltonian that refers
to the spin only. And since $\left\vert \mathbf{B}\right\vert $ is connected
with the intensity of the field~(\ref{omega}), for large $n$ we
write\footnote{The explicit form of relation~(\ref{omega}) depends on the
ordering definition for $|\mathbf{\hat{b}|}^{2}$. Semiclassical and fully
quantized quantities can only be compared in a limit where ordering is not
relevant.}
\begin{equation}
\hat{H}_{1}^{\left(  n\right)  }=\frac{\delta}{2}\sigma_{3}+B\sigma
_{1}~,\label{hn}%
\end{equation}
which is the classical Hamiltonian~(\ref{hlinha}) in the rotating picture.

It should be pointed out that, although the decomposition in Eq.~(\ref{h3}) is
well known, our treatment clarifies the relation between the results in
Schr\"{o}dinger picture with the semiclassical limit of the quantum theory.
For this purpose, the interpretation of the operator $\exp(i\omega\hat{N}t)$
as a rotation is essential. Besides, the semiclassical limit must be taken in
the Schr\"{o}dinger picture because, since the Hamiltonian is time independent
in this representation, only in the rotating frame the (semiclassical)
Schr\"{o}dinger and Heisenberg Hamiltonians coincide.

\subsection{Relation with the Jaynes-Cummings model}

In the present section, we will establish connections between the
Jaynes-Cumming model, that involves a second quantized treatment of a linearly
polarized field, and the original Rabi setup that presents a semiclassical
approach involving a circularly polarized field.

Let us consider a linear field,
\begin{equation}
\mathbf{\tilde{B}}=\left(  2B\cos\omega t,0,0\right)  \,\,.
\end{equation}
In this case, since the field intensity is not constant, we can not discard
the electric field in the Hamiltonian. Still, this problem can be fixed if we
define
\begin{equation}
c=\sqrt{\frac{1}{2\omega}}\left(  \tilde{B}_{1}-iE_{2}\right)  ~,
\end{equation}
where we use that, from the Maxwell's equations in vacuum, $\mathbf{E}$ and
$\mathbf{\tilde{B}}$ are orthogonal, with a phase difference of $\pi/2$. 
It follows that
\begin{equation}
H_{f}^{\prime}=\frac{1}{2}\left(  B^{2}+E^{2}\right)  =\omega c^{\ast
}c~.\label{ljc}%
\end{equation}
But, since the spin interacts only with the magnetic field, we obtain that
\begin{equation}
\hat{H}_{I}=\tilde{B}_{1}\sigma_{1}=\sqrt{\frac{\omega}{2}}\left(  c+c^{\ast
}\right)  \sigma_{1}\rightarrow\sqrt{\frac{\omega}{2}}\hat{B}\hat{S}%
~,\ \hat{B}=\hat{c}+\hat{c}^{+}~,\label{ijc}%
\end{equation}
where $\hat{S}$ is the polarization operator introduced in Eq.~(\ref{jc}).

The problem can be simplified writing the interaction Hamiltonian in the form
(\ref{hl0}), that is $\hat{H}_{I}=\hat{H}_{+}+\hat{H}_{-}$. In Dirac
notation,
\begin{equation}
\hat{H}_{\pm}=\left(  \boldsymbol{\sigma}\mathbf{B}_{\pm}\right)  =B\left[
e^{\mp i\omega t}\left\vert +\right\rangle \left\langle -\right\vert +e^{\pm
i\omega t}\left\vert -\right\rangle \left\langle +\right\vert \right]  ~.
\label{hl-1}%
\end{equation}
Previous expression shows that it is possible to apply the second quantization
scheme to theories equipped with the Hamiltonians $\hat{H}_{\pm}$, using the
quantity $\sqrt{\omega/2}b=\left(  B_{1}-iB_{2}\right)  =Be^{-i\omega t}$
defined in Eq.~(\ref{hc}) for the circular field $\mathbf{B}$. It should be
observed that
\begin{equation}
\tilde{B}_{1}=\sqrt{\frac{\omega}{2}}\left(  b+b^{\ast}\right)  =\sqrt
{\frac{\omega}{2}}\left(  c+c^{\ast}\right)  ~,
\end{equation}
and so the field operator $\hat{B}$ (\ref{ijc}) has the same form when written
in terms of $\hat{b}$ or $\hat{c}$, that is, $\hat{B}=\hat{c}+\hat{c}^{+}%
=\hat{b}+\hat{b}^{+}$. It follows that, even working with $b$ in the
interaction Hamiltonian, it is still possible to use the same free field
Hamiltonian~(\ref{ljc}). The Jaynes-Cummings model represents the second
quantization of this hybrid description.

The second quantization of the Hamiltonians in Eq.~(\ref{hl-1}) gives
\begin{equation}
\hat{H}_{\pm}=\sqrt{\frac{\omega}{2}}\left(  \hat{b}\left(  t\right)
\left\vert \pm\right\rangle \left\langle \mp\right\vert +\hat{b}^{+}\left(
t\right)  \left\vert \mp\right\rangle \left\langle \pm\right\vert \right)  ~.
\label{h--1}%
\end{equation}
The interaction Hamiltonian in this system is
\begin{equation}
\hat{H}_{I}=\hat{H}_{+}+\hat{H}_{-}=\sqrt{\frac{\omega}{2}}\hat{B}\hat
{S}~,\ \hat{S}=\left\vert +\right\rangle \left\langle -\right\vert +\left\vert
-\right\rangle \left\langle +\right\vert ~, \label{hjc-1}%
\end{equation}
which (for $\mu=-1$) is the same original Hamiltonian~(\ref{ijc}). This was
the interaction Hamiltonian suggested by Jaynes-Cummings \cite{jc}. The fact
that
\begin{equation}
\hat{H}_{f}^{\prime}=\omega\hat{c}^{+}\hat{c}\neq\omega\hat{b}^{+}\hat{b}%
\end{equation}
is the cause of the apparent instantaneous non conservation of energy in the
Jaynes-Cummings model due to the counter-rotating terms.\footnote{This term
has the form $\hat{b}^{+}\left\vert +\right\rangle \left\langle -\right\vert
$, representing the spin excitation and the simultaneous photon emission. In
our description, they promote transitions between sectors described by
$\hat{H}^{\left(  n\right)  }$ with different $n$.} In other words, $\hat
{b}^{+}\hat{b}$ is not the number operator of the problem.

As it is well known, focusing on the interaction Hamiltonian only, the
Jaynes-Cummings model complemented with the rotating wave approximation is
equivalent to the Rabi model with circular field. However, it should be
observed that, in this approximation, $\hat{H}_{I}\simeq\hat{H}_{+}$ and
$\hat{c}+\hat{c}^{+}\neq\hat{b}+\hat{b}^{+}$. Therefore, the free field
Hamiltonian is not given by~(\ref{ljc}). This discrepancy does not change
the transition probability between states near the resonance frequency. But
must be taken into account if, for example, the interaction picture is used in
models far from resonance point.

\section{\label{spinorial-field}Quantization of the spinorial field}

The second quantization procedure can be extended to all fields in the Rabi
problem, including the spinorial field. Here, we will consider the Pauli
spinorial field and a non-covariant formulation.\footnote{The approach here
differs from many usual physics treatments of spinor quantization, where the
starting point is a relativistic wave equation.} In this development, some
attention should be given to possible failures of a classical symmetry in its
quantized version. As it is well known, anomalies are a common feature of
second quantized models. In particular, the relation between the spin and the
Bloch sphere (i.e., the $SU(2)$ symmetry) is not necessarily maintained in a
second quantized approach.

Following the work in the previous section, we will construct a Hamiltonian
for the free spinorial field. In this case, the presence of the $B_{3}$ term
is associated to the energy content of the system, and therefore we consider a
free field when the field component in the plane $x\times y$ is null. The free
spin Hamiltonian is
\begin{equation}
\hat{H}_{\mathrm{sf}}=B_{3}\sigma_{3}~.
\end{equation}
We can associate the measured energy of the system with the average value of
the energy. Hence the energy of the free spinorial field is
\begin{equation}
H_{\mathrm{sf}}=\left\langle \psi\right\vert \hat{H}_{\mathrm{sf}}\left\vert
\psi\right\rangle =B_{3}\left(  \left\vert \psi_{1}\right\vert ^{2}-\left\vert
\psi_{2}\right\vert ^{2}\right)  ~, \label{hsf}%
\end{equation}
where $\{\psi_{i}\}$ are the components of the Pauli spinor $\left\vert
\psi\right\rangle $. This Hamiltonian can be written in the form
\begin{equation}
H=\sum_{i}\alpha_{i}b_{i}^{\ast}b_{i}~,\ \alpha_{i}\in\mathbb{R}~, \label{hsq}%
\end{equation}
with the identification $b_{i}=\psi_{i}$. Since we are treating fermions, the
spinors obey the anti-commutation rules
\begin{equation}
\left\{  \hat{\psi}_{i},\hat{\psi}_{j}^{+}\right\}  =\delta_{ij}~,\ \left\{
\hat{\psi}_{i},\hat{\psi}_{j}\right\}  =\left\{  \hat{\psi}_{i}^{+},\hat{\psi
}_{j}^{+}\right\}  =0~. \label{anti}%
\end{equation}
Eq.~(\ref{hsf}) is now written as
\begin{equation}
\hat{H}_{\mathrm{sf}}=B_{3}\left(  \hat{\psi}_{1}^{+}\hat{\psi}_{1}-\hat{\psi
}_{2}^{+}\hat{\psi}_{2}\right)  ~.
\end{equation}

In the present description, for $\omega B_{3}>0$, the operator $\hat{\psi}%
_{1}$ acts as an annihilation operator of the positive energy states and as an
identity operator in the space of negative energy states. An analogous
procedure is applied to the operator $\hat{\psi}_{2}$. After the quantization
of the magnetic and Pauli spinorial field, we have
\begin{equation}
\hat{H}_{sI}=\Omega\left(  \hat{\psi}_{1}^{+}\hat{\psi}_{2}\hat{b}+\hat{\psi
}_{1}\hat{\psi}_{2}^{+}\hat{b}^{+}\right)  ~. \label{hiq}%
\end{equation}

We stress that in our approach, unlike Pauli's theory, the definition of which
state $\{\left\vert +\right\rangle ,\left\vert -\right\rangle \}$ has positive
energy is not determined only by the sign of $B_{3}$, but also by $\omega$.
This happens because the quantization of the Maxwell field was developed in
such a way that the photons always have positive energy (given by $\left\vert
\omega\right\vert $). It follows that, for $\omega<0$, we should exchange
$\hat{b}\leftrightarrow\hat{b}^{+}$ in Eq.~(\ref{hiq}). For $B_{3}>0$,
$\hat{\psi}_{1}^{+}$ ($\hat{\psi}_{2}^{+}$) creates particles in the positive
sector with $\omega>0$ ($\omega<0$).

From this latter observation, we obtain an interpretation for the fact that
only photons associated to the rotating term are capable to promote the
excitation of the spinning particle near the resonance point. Notice that this
happens despite the fact that photons associated to both rotating and
counter-rotating terms have the same energy. The point is that, while in the
semiclassical limit this is a consequence of the conservation of angular
momentum, in the full quantum theory the counter-rotating photons are not in
the adequate energy sector.

We now consider the semiclassical limit of the spinorial field. This limit is
given by the matrix elements of the Hamiltonian in a suitable basis, the
physical states of the system $\left\{  \left\vert 0,1\right\rangle
,\left\vert 1,0\right\rangle \right\}  $. This basis is chosen since its
elements are the only states coupled by the interaction Hamiltonian~(\ref{hiq}%
).
Let us introduce the following operators \cite{sch}:
\begin{equation}
\hat{J}_{+}=\hat{\psi}_{1}^{+}\hat{\psi}_{2}~,\ \hat{J}_{-}=\hat{\psi}_{2}%
^{+}\hat{\psi}_{1}~,\ \hat{J}_{3}=\frac{1}{2}\left(  \hat{\psi}_{1}^{+}%
\hat{\psi}_{1}-\hat{\psi}_{2}^{+}\hat{\psi}_{2}\right)  ~.\label{sw}%
\end{equation}
With the elements presented in Eq.~(\ref{sw}), the interaction Hamiltonian is
written as
\begin{equation}
\hat{H}_{sI}=\Omega\left(  \hat{J}_{+}\hat{b}+\hat{J}_{-}\hat{b}^{+}\right)
~.\label{hi2}%
\end{equation}

The important aspect in the definitions of $\hat{J}_{\pm}$ and $\hat{J}_{3}$
is that, using the anti-commuting relation rule in Eq.~(\ref{anti}), we
observe that the operators $\hat{J}_{\pm}$ and $\hat{J}_{3}$ satisfy the
relations
\begin{equation}
\left[  \hat{J}_{3},\hat{J}_{\pm}\right]  =\pm\hat{J}_{\pm}~,\ \left[  \hat
{J}_{+},\hat{J}_{-}\right]  =2\hat{J}_{3}~,
\end{equation}
which is the algebra of the angular momentum operators,
\begin{equation}
\hat{J}_{+}=\hat{J}_{1}+i\hat{J}_{2}=\hat{J}_{-}^{+}~,\ \left[  \hat{J}%
_{i},\hat{J}_{k}\right]  =i\varepsilon_{ijk}\hat{J}_{k}~.
\end{equation}
That is, the operator algebra generated by $\{\hat{J}_{i}\}$ can be satisfied
setting $\hat{J}_{i}=\sigma_{i}/2$~. The same conclusion can be obtained by
the direct calculation of the matrix elements of $\hat{J}_{\pm}$ in
Eq.~(\ref{sw}) with the basis of physical states $\{\left\vert -\right\rangle
\equiv\left\vert 0,1\right\rangle ,~\left\vert +\right\rangle \equiv\left\vert
1,0\right\rangle \}$. In this basis
\begin{equation}
\hat{J}_{+}=\left\vert +\right\rangle \left\langle -\right\vert =\frac{1}%
{2}\left(  \sigma_{1}+i\sigma_{2}\right)  ~,\ \hat{J}_{-}=\left(  \hat{J}%
_{+}\right)  ^{+}=\left\vert -\right\rangle \left\langle +\right\vert ~.
\end{equation}
Using this representation, we can write Eqs.~(\ref{hsf}) and (\ref{hi2}) as
\begin{equation}
\hat{H}_{\mathrm{sf}}=B_{3}\sigma_{3}~,\ \hat{H}_{sI}=\Omega\hat{b}\left\vert
+\right\rangle \left\langle -\right\vert +\Omega\hat{b}^{+}\left\vert
-\right\rangle \left\langle +\right\vert ~. \label{HsI}%
\end{equation}
The operators $\hat{H}_{\mathrm{sf}}$ and $\hat{H}_{sI}$ in Eq.~(\ref{HsI})
are the components associated to the spin and to the interaction in the
Hamiltonian~(\ref{hs}). Therefore, the Hamiltonian obtained with the second
quantization approach in this work has the correct semiclassical limit.

\section{\label{final-comments}Final comments}

In the present work, a second quantization approach is developed for the Rabi
Problem. In this setting, we explicitly show that a non-covariant theory with
interaction can be directly quantized. In our development, the quantization is
made without the rotating wave approximation, and therefore our results are
not equivalent to the Jaynes-Cumming approach. We also show how Pauli
spinorial field can be directly quantized in an important (although
particular) setup. Moreover, in our work a ``complete field'' approach is
developed, where both electromagnetic and matter sectors are quantized. The
appearance of a negative energy sector for the spinorial field implies a
specific identification between Schr\"{o}dinger's and Maxwell's theories. In
the development reported here, it was possible to establish a connection
between semiclassical and purely quantum physical quantities.

Unlike the usual development for the electromagnetic field, the procedure
presented here gives a simple and direct method to quantize the magnetic
field, in a concrete scenario. We stress that, besides being simpler, the
direct quantization of the electromagnetic field (instead of the usual
quantization of the electromagnetic potential) can avoid problems. For
instance, as already observed by Lamb \cite{Lamb52}, a perturbative treatment
of a dipole transition based on the electromagnetic potential (and not on the
electromagnetic field) can result in spurious terms and a significant
distortion in the resonant curve of the system.

One important point addressed is the fact that the classical quantities in the
right side of the Heisenberg equation~(\ref{emh}) should be associated to an
inertial frame, when non-covariant theories are considered. Only then the
semiclassical limit of the quantum theory furnishes the original classical
model. Particularly for the Rabi problem, despite the quantization obtained in
both Schr\"{o}dinger and Heisenberg pictures are unitarily equivalent (as
expected), the procedure is only consistent if the rotating frame is
associated to the Schr\"{o}dinger picture. As done in the usual quantization
of covariant field theories, where vacuum is chosen using Lorentz symmetry as
criteria, in our non-covariant approach the correct quantization is chosen
using as criteria the presence of classical symmetries in the final quantum
theory. This point can be interpreted as a correspondence principle in the
second quantization approach. In fact, the development here would not be
possible without the explicit use of the mentioned correspondence principle.
This remark is very important, but it appears not to have been fully
considered in the pertinent literature.

The fact that it is possible to define in the Rabi problem a number operator
which is time independent is a consequence of the constancy of the magnetic
field magnitude. This is the reason why we could disregard the electric field
and still obtain a Hamiltonian where the energy is conserved. This would not
be true if a linearly polarized field were considered. But even in this latter
case the development presented here could be applied. The work developed here
explicits the similarities (and differences) between the Rabi problem and the
analogous scenario with a linearly polarized field.

\newpage

The present work also clarifies the difference between the Rabi problem and
the Jaynes-Cummings model with the rotating wave approximation. That is,
although the interaction Hamiltonians in both cases coincide, the free field
Hamiltonians remain distinct. Still, it is possible to define a consistent
number operator for the Jaynes-Cummings model, which can be used in the
construction of the interaction picture and in the treatment of transitions
with high detuning.

Finally, we expect that the main features of the approach might be relevant
even when more general non-covariant theories are considered. For example, in
the treatment of semiclassical models that use exact solutions of the spin
equation, such as the adiabatic magnetic pulse \cite{BurLoDS99}. This setup is
potentially important in the manipulation of quantum dots, and should be
considered in a future development of the present work.

\begin{acknowledgments}
C. M. is supported by FAPESP, Brazil [grant number 2015/24380-2]; and
CNPq, Brazil [grant number \linebreak 307709/2015-9].
\end{acknowledgments}

\end{document}